\documentclass[letterpaper,twocolumn,10pt]{article}
\usepackage{usenix2024_SOUPS}
\usepackage{enumitem}
\setlist{nosep}
    
\usepackage{tikz}
\usepackage{chngcntr}
\usepackage{amsmath}
\vspace{-1em}
\usepackage{authblk}
\begin{document}

\title{\Large \bf User Awareness and Perspectives Survey on Privacy, Security and Usability of Auditory Prostheses}

\def\plainauthor{Author name(s) for PDF metadata. Don't forget to anonymize for submission!}

\author[1]{Sohini Saha}
\author[1]{Leslie M. Collins}
\author[2]{Sherri L. Smith}
\author[1]{Boyla O. Mainsah}
\affil[1]{Department of Electrical and Computer Engineering, Duke University}
\affil[2]{Department of Head and Neck Surgery \& Communication Sciences, Duke University Medical Center}

\vspace{-3em}
\maketitle
\vspace{-3em}
\thecopyright
\vspace{-3em}
\section{Motivation}
\vspace{-1em}
According to the World Health Organization, over 466 million people worldwide suffer from disabling hearing loss, with approximately 34 million of these being children \cite{WHO2024Deafness}. Auditory prosthetic devices, such as hearing aids (HA) and cochlear implants (CI), have become indispensable tools for restoring hearing and enhancing the quality of life for individuals with hearing impairments \cite{Carlyon2021CochlearIR,Hoppe}.  

The adoption of HA and CIs has been on the rise. Clinical research and consumer studies indicate that users of HA and CIs report significant improvements in their daily lives, including enhanced communication abilities and social engagement and reduced psychological stress \cite{Chisolm2007Systematic,Kabis2023Challenges,PMID35774034,Marinelli, MarcosAlonso2023FactorsIT, Cox2016ImpactOH}. Other studies focusing on stress evaluation have shown that auditory prostheses can considerably alleviate the mental strain associated with hearing loss, thus contributing to better mental health outcomes for users \cite{Kamil2015TheEO, Mukherjee, Saki2023TheIO}. Accessibility remains a critical area of focus, with efforts aimed at making these devices more affordable and widely available, particularly in low- and middle-income countries \cite{Bright2018ASR,Goulios, Olusanya}.

Modern auditory prosthetic devices are more advanced and interconnected with digital networks to add functionality, such as streaming audio directly from smartphones and other devices, remote adjustments by audiologists, integration with smart home systems, and access to artificial intelligence-driven sound enhancement features \cite{s24051546, Lesica, encyclopedia4010011}. With this interconnectivity, issues surrounding data privacy and security have become increasingly pertinent. Despite the growing relevance in the domain of auditory prostheses, there is a notable gap in the literature regarding consumer awareness of privacy and security concerns related to these technologies. Studies have highlighted the lack of cybersecurity in private medical practices in audiology, indicating increased risks of cyberattacks and security breaches involving sensitive personal information in CI and HA devices \cite{Dykstra2020CybersecurityIM, Srdjan2010OnTS, Katrakazas2018AO}.   This underscores the critical need for consumer awareness of potential vulnerabilities. 

There is limited research on the usability perceptions of current HA and CI models from the perspective of end-users. Usability is directly linked to privacy and security because if users find the devices complex or unintuitive, they may not effectively manage privacy settings or recognize security threats \cite{info14120641}. Therefore, enhancing usability can improve consumer awareness and practices related to privacy and security. This underscores the importance of understanding user experiences and concerns. 

In addition, no studies have investigated consumer mental models during the purchasing process, particularly which factors they prioritize when selecting a device. Understanding these factors is crucial for ensuring that the devices not only meet auditory needs but also align with the expectations and concerns of the users.   By connecting these aspects—technological advancement, consumer awareness, usability, and purchasing behavior—we can better address the comprehensive needs of end-users, ensuring their safety, satisfaction, and overall well-being.

To this end, our work-in-progress study aims to gain insight into consumer perceptions and awareness concerning privacy and security matters associated with cochlear implants and hearing aids by answering the following research questions:

\begin{itemize}
 \item[RQ1] How satisfied are consumers with their current cochlear implants or hearing aids?
\item[RQ2] What is the perceived level of importance of different factors like performance, durability, and price while making a purchase decision?
\item[RQ3] How informed are users about privacy and security practices?
\item[RQ4] What strategies might be helpful to increase consumer awareness about privacy and security features?
\end{itemize}

This work contributes to the field by addressing gaps in user perceptions of HA and CI usability, identifying key factors in consumer purchasing decisions, and highlighting the need for improved privacy and security awareness and education among users.
\vspace{-1em}
\section{Methodology}
\vspace{-1em}
We developed a survey on the Research Electronic Data Capture (REDCap) platform \cite{Harris2009ResearchED, Harris}. The survey encompassed 45 questions  grouped in multiple sections covering demographics, device details, usage satisfaction, mobile app security, privacy, cybersecurity awareness, improvements, and purchase decisions. The survey includes Likert scale questions and open-ended questions.

The survey study was approved by the Duke University Institutional Review Board. We recruited participants by posting advertisements and flyers in private social media groups dedicated to CI/HA users. The participants were compensated 10 USD for successfully completing the survey. We recruited 27 participants, comprising 12 cochlear implant users and 15 hearing aid users. Participant demographics are summarized in Appendix Figure \ref{fig:demographics}. Details about participants’ brand of auditory prosthetic device and duration of device usage are summarized in Appendix Figure \ref{fig:device}.


\vspace{-1em}
\section{Results}
\begin{figure*}
\centering
\includegraphics[scale = 0.5]{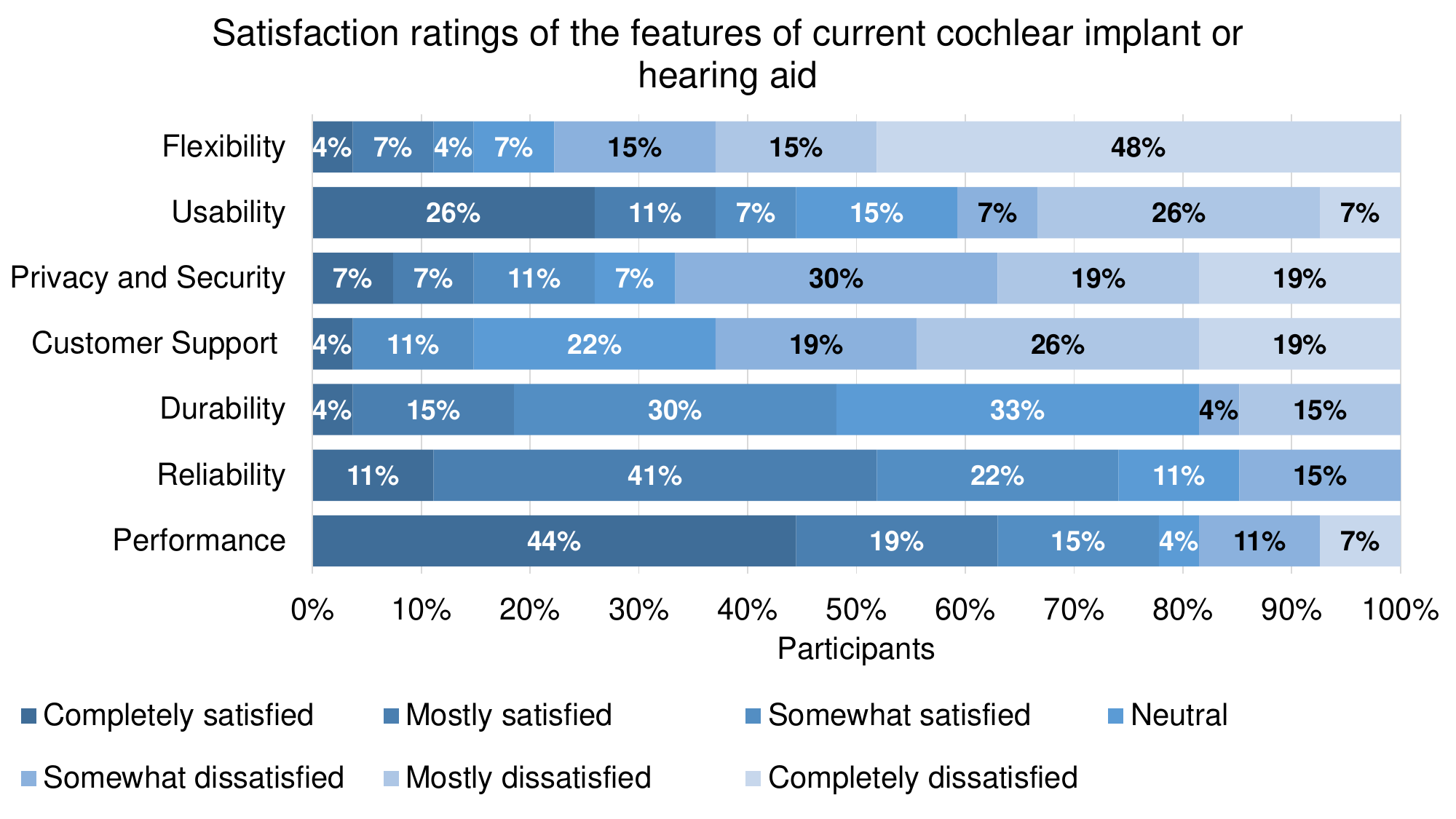}
\caption{\label{fig:satisfaction} Survey participants’ responses when asked: \textit{How satisfied are you with the features of your current cochlear implant or hearing aid?}}
\end{figure*}

\vspace{-1em}
A subset of participant survey results is reported here. We assessed participants' satisfaction levels on the importance of various features of their auditory prosthesis. The device features included: \textit{performance} of the CI/HA in improving hearing and communication; \textit{reliability} in consistent functioning; \textit{durability} of the device to withstand wear and tear; \textit{customer support} from the manufacturer, \textit{privacy and security} of personal data; \textit{usability} in everyday life; and \textit{flexibility} in the number device settings based on the listening environment. 44\% of participants reported complete satisfaction with performance, while 48\% expressed complete dissatisfaction with flexibility (Appendix Figure \ref{fig:satisfaction}) (RQ1). 

\begin{figure*}
\centering
\includegraphics[scale = 0.5 ]{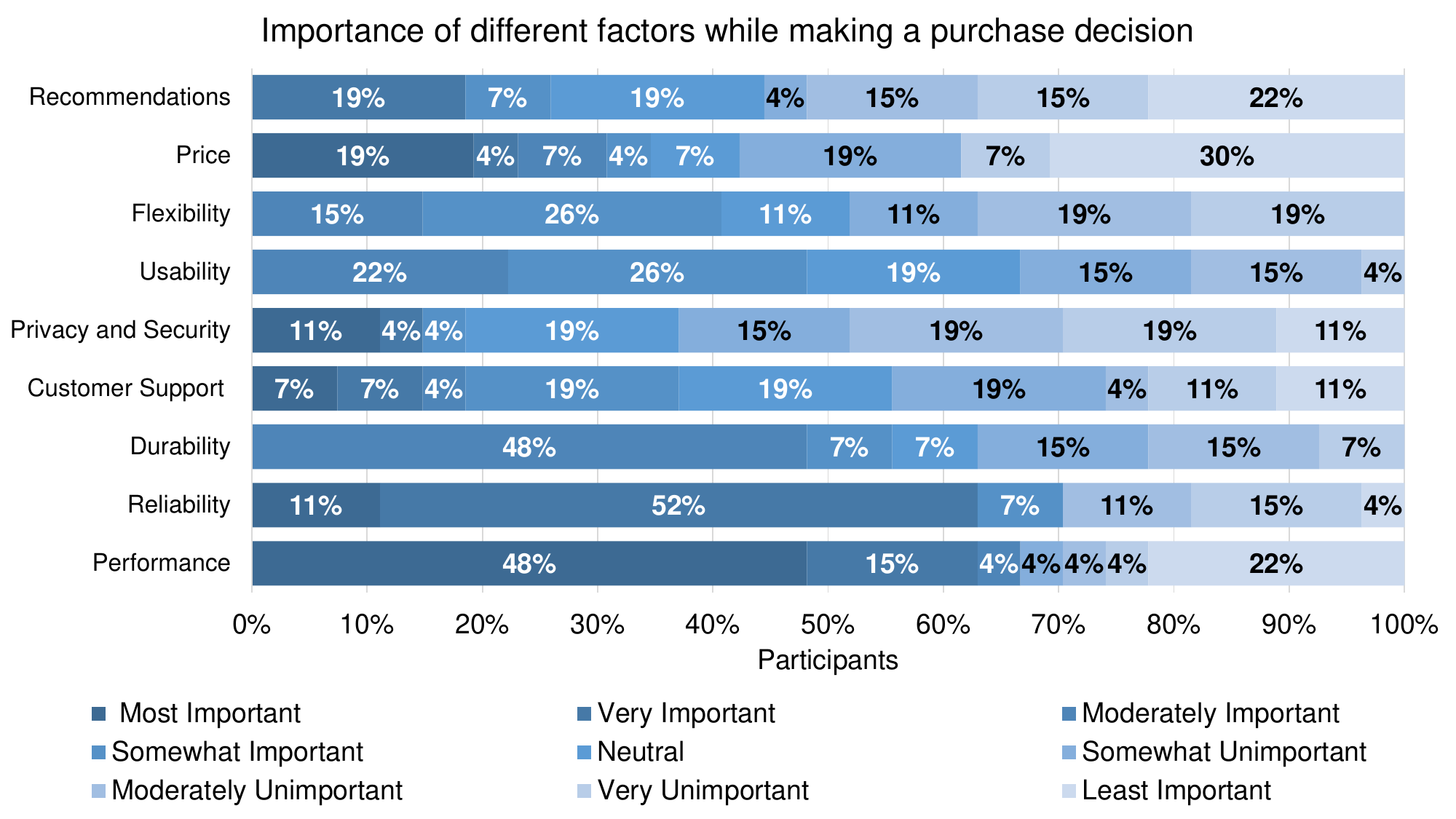}
\caption{\label{fig:importance} Survey participants’ responses when asked: \textit{When deciding to purchase a cochlear implant or hearing aid, what factors are most important to you? Please rank them in order of importance, with 1 being the most important and 9 being the least important. Please assign a unique rank to each option.}}
\end{figure*}

Participants were also asked to rank the device’s features based on their perceived level of importance when deciding which device to purchase (RQ2).  Comparable to satisfaction levels, 48\% of participants considered performance to be the most important factor when making a purchase decision. Reliability (52\%), durability (48\%) and usability (22\%) were the next highly ranked factors. Interestingly, price (30\%) and recommendations from healthcare professionals (22\%) were ranked the least important factors by most participants, while privacy, security, and customer support garnered mostly neutral sentiments (Appendix Figure \ref{fig:importance}).

\begin{figure}
\centering
\includegraphics[scale = 0.2]{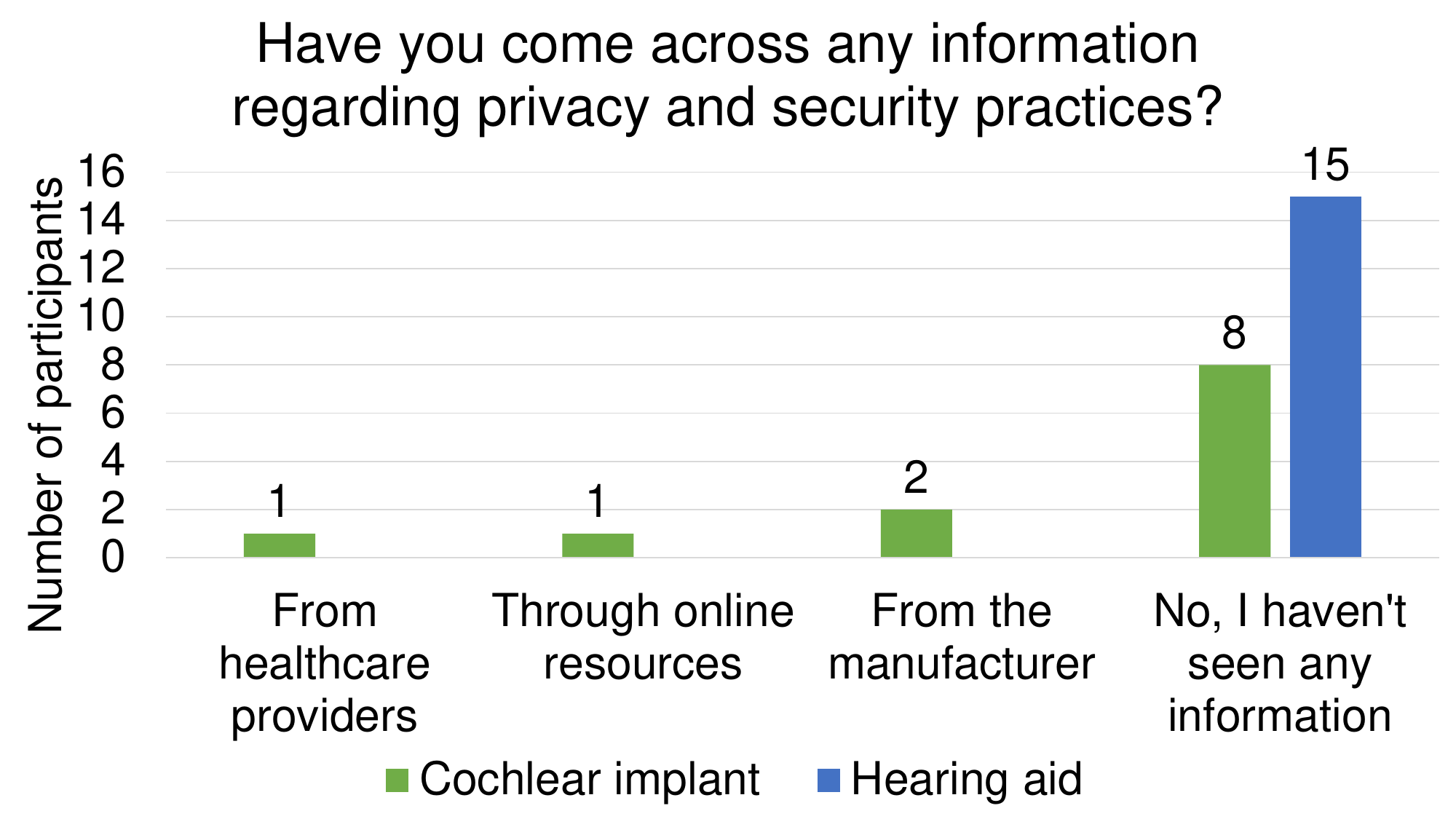}
\includegraphics[scale = 0.2]{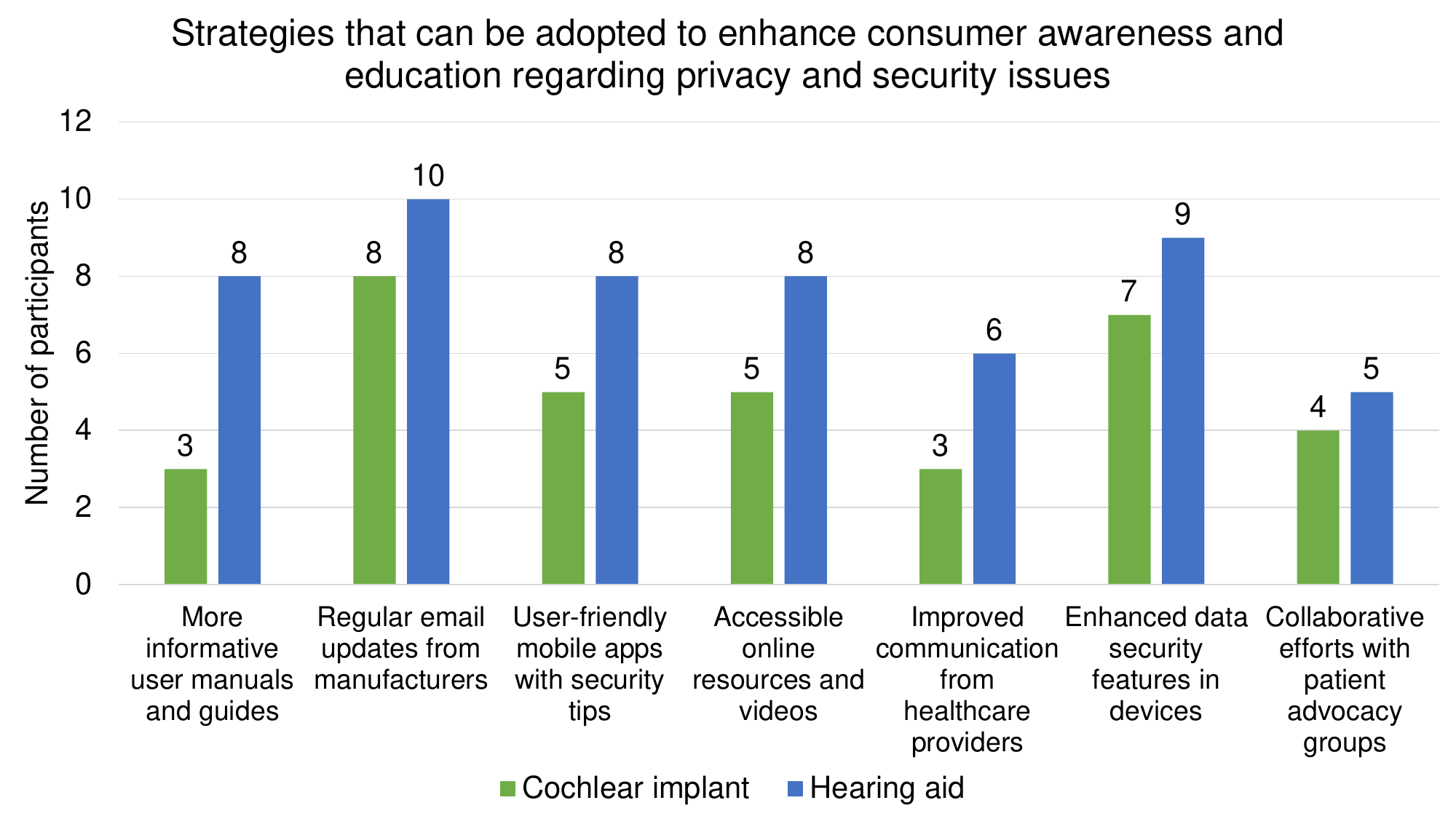}
\caption{\label{fig:privacyandsecurity} Participants’ responses when asked questions on device privacy and security: \textit{Have you come across any information regarding privacy and security practices (like usage of passwords, and information on data protection) associated with your cochlear implant or hearing aid?} (top) and \textit{What do you believe could be done to enhance consumer awareness and education regarding privacy and security issues associated with cochlear implants or hearing aids?} (bottom)}
\end{figure}

Most participants (23 out of 27) were found to be uninformed about privacy and security practices (such as password usage and data privacy) associated with the devices, Figure \ref{fig:privacyandsecurity} (RQ3). When queried on strategies that could be adopted to enhance user awareness and education on privacy and security issues related to their devices, the most common responses were receiving regular email updates from manufacturers (18/27) and enhanced data security features in their devices (16/27).   Other practices that participants selected included accessibility to online resources and videos (13 out of 27), availability of user-friendly mobile apps and security tips (13 out of 27), more informative user manuals and guides (11 out of 27), improved communication from healthcare providers (9 out of 27), and collaborative efforts with patient advocacy groups (9 out of 27), Figure \ref{fig:privacyandsecurity} (RQ4).  

\vspace{-1em}
\label{sec:figs}

\section{Discussion}
\vspace{-1em}
Our study highlights that users of CI and HA devices express a strong satisfaction with their device performance in enhancing their hearing and communication abilities, which significantly influences their decision-making process when purchasing these devices. However, a significant gap in awareness regarding privacy and security practices was evident. Participants  expressed interest in obtaining guidance to augment their understanding of privacy and security matters, emphasizing the need for comprehensive knowledge dissemination. Our findings underscore the importance of bolstering privacy and security settings while ensuring user-friendliness, ultimately contributing to the safety and satisfaction of CI/HA device users. Given the increasing integration of wireless technology into CI/HA devices, enhancing user awareness about privacy and security issues related to their auditory prostheses is important.
Future work could involve developing educational programs to enhance user knowledge of privacy and security practices, investigating the impact of this awareness on device adoption and usage, and designing more secure and user-friendly features for HAs and CIs. Additionally, exploring the long-term effects of improved privacy measures on user satisfaction and potential regulatory implications could foster valuable collaborations.
\section*{Acknowledgments}
\vspace{-1em}
This work was supported by the National Institutes of Health (NIH) grant 1R56DC020267-01A1. The support for the Duke Office of Clinical Research to host REDCap is made possible by grant UL1TR001117 from the National Center for Research Resources, a component of the NIH and NIH Roadmap for Medical Research.

\bibliographystyle{unsrt}
\bibliography{usenix2024_SOUPS}

\appendix

\counterwithin{figure}{section}
\section{Appendix}
\newpage
\begin{figure}
\centering
\includegraphics[scale = 0.3]{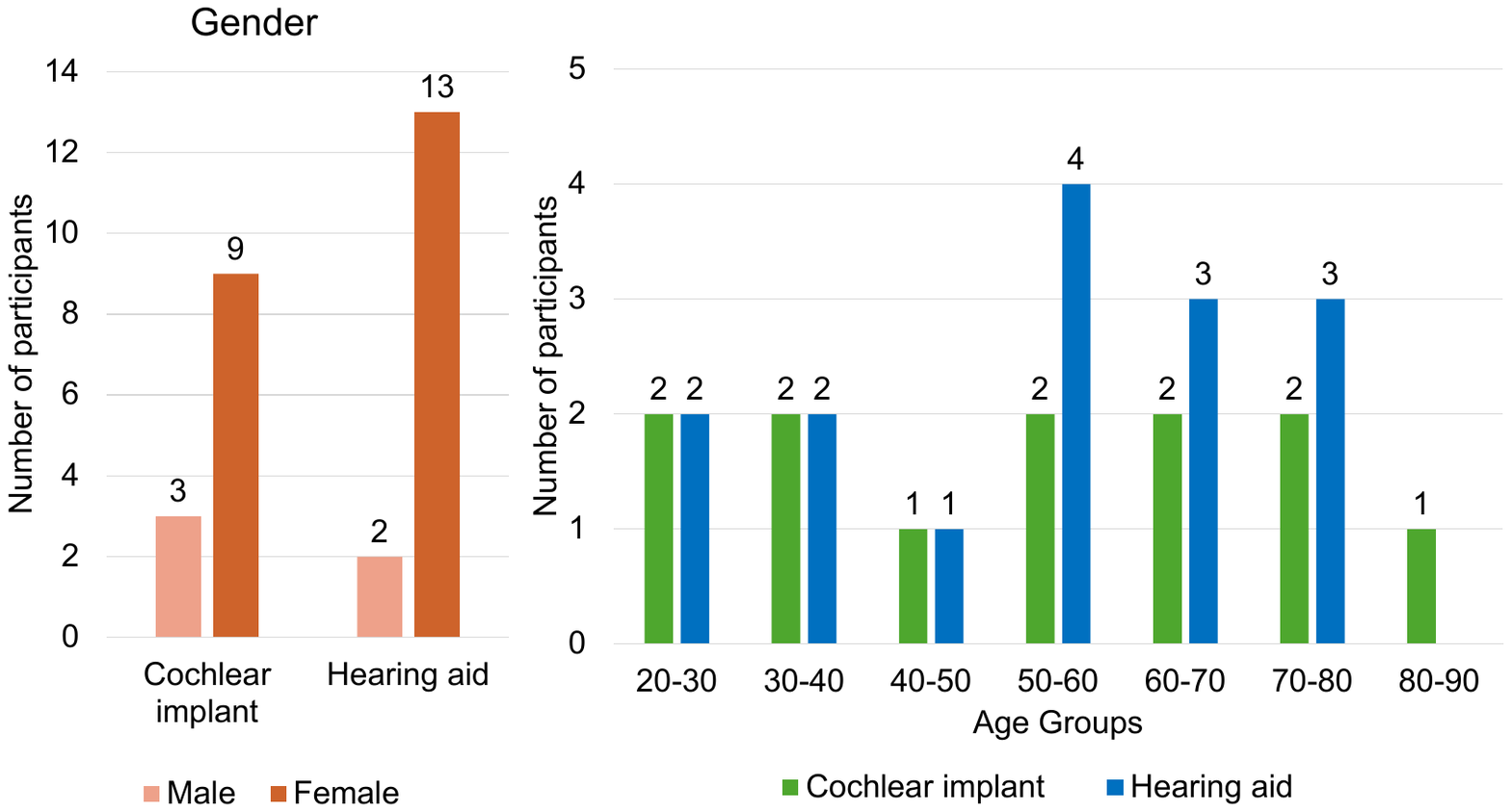}
\caption{\label{fig:demographics} Participants demographics. Gender (left) and age-groups of the participants (right).}
\end{figure}

\begin{figure}
\centering
\includegraphics[scale = 0.3]{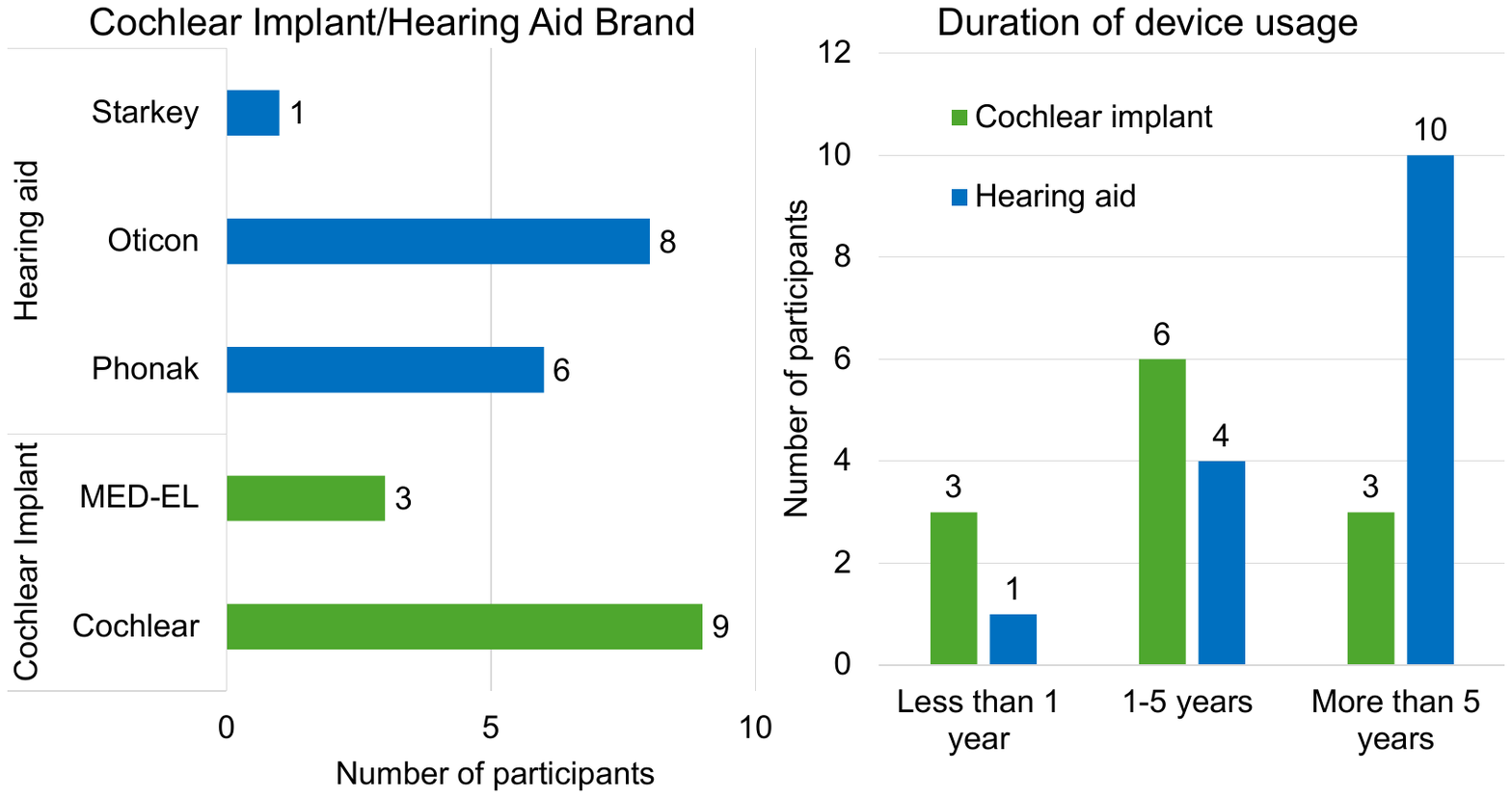}
\caption{\label{fig:device} Distribution of participants across various brands of cochlear implants and hearing aid devices (left) and duration of device usage among participants (right).}
\end{figure}

\end{document}